# A heavy-tailed and overdispersed collective risk model

Pamela M. Chiroque-Solano[1] and Fernando Antônio da S. Moura[2]

[1] *Department of Statistics, Federal University of Rio de Janeiro, Brazil, Br, email:pamela@dme.ufrj.br*

[2] *Department of Statistics, Federal University of Rio de Janeiro, Brazil, Br, email:fmoura@im.ufrj.br*

ABSTRACT

Insurance data can be asymmetric with heavy tails, causing inadequate adjustments of the usually applied models. To deal with this issue, hierarchical models for collective risk with heavy-tails of the claims distributions that take also into account overdispersion of the number of claims are proposed. In particular, the distribution of the logarithm of the aggregate value of claims is assumed to follow a Student-t distribution. Additionally, to incorporate possible overdispersion, the number of claims is modeled as having a negative binomial distribution.

Bayesian decision theory is invoked to calculate the fair premium based on the modified absolute deviation utility. An application to a health insurance dataset is presented together with some diagnostic measures to identify excess variability. The variability measures are analyzed using the marginal posterior predictive distribution of the premiums according to some competitive models. Finally, a simulation study is carried out to assess the predictive capability of the model and the adequacy of the Bayesian estimation procedure.

## 1. Introduction

The presence of observations far from the center of the data often occurs in insurance data sets. This can be the result of the presence of outliers in the dataset or a heavy-tailed distribution. In the actuary context, many authors have proposed classical approaches to deal with both issues. Under the frequentist approach, the proposal of using robust estimators to attenuate the impact of outliers was first introduced by Kunsch (1992). He examined cases of claims with extreme values and proposed the use of an M-robust estimator instead of the linear estimator of credibility.

In this context, Gisler and Reinhard (1993) proposed the use of the T-robust estimator. They used a particular influence function that reduced the influence of observations larger than a certain pre-specified value.

Klugman and Hogg (1983) presented an empirical analysis of the size of loss in casualty insurance data and observed that these kinds of data have a very heavy-tailed and skewed distribution. On the other hand, there are several works describing how to model heavy-tailed distributions and detect outliers using the Bayesian approach. West (1984) considered a robust regression model to fit heavy-tailed data. The distribution assigned to the error term was structured as a mixture in the scale parameter of the normal distribution. In the actuarial context, Cipra (1996) included outliers and missing observations in a dynamic credibility model. Huisman et al. (1998) applied heavy-tailed distributions to FX return. Further details on how models for return response with outliers are fitted can be seen in Günay (2017). Bühlmann and Gisler (2005, Chapter 2) presented the Bayes premium under the principles of credibility theory. These concepts are highly interpretable and understandable in actuary practice.

Based on the collective risk model and the principles of credibility theory, our challenge is to obtain robust estimates in scenarios where heavy-tailed distributions are fitted under the Bayesian paradigm. The most popular form in the literature is that the distribution assigned to the error term is structured as a mixture in the scale parameter of the normal distribution. This is obtained through a scale mixture of Gaussian and inverse-gamma distributions and is equivalent to considering all of the Gaussian errors as Student-t errors. For a comprehensive overview of other types of mixture models commonly used to detect and account for outliers, see Hyungsuk et al. (2019).

In the actuarial context, the log Student-t distribution with a known degree of freedom was first proposed by Klugman and Hogg (1983).

Our contribution here is focused on incorporating a heavy tail to the response error in the values of claims belonging to the collective risk process. Other possibilities to incorporate robustness in the estimation can be seen in Andrade and O'Hagan (2011) and Desgagné and Gagnon (2019), among others.

In this direction, our principal contribution is to propose a collective risk model in the Bayesian framework that accommodates the presence of observations far from the center of the data. We propose to model the logarithm of the claim values as Student-t distributed with a degree of freedom parameter assumed to be unknown. Another common aspect of the uncertainty inherent to insurance data is the overdispersion issue. To take into account possible overdispersion of the claim number distribution, we replace the Poisson distribution with the negative binomial one. We also analyze the parameter that quantifies the extra-Poisson variation (Zhou et al., 2012). To identify groups of risks, two measures are calculated: $VaR$ (Value at Risk) and $TVaR$ (Tail Value at Risk). These measures began being used by insurance companies to assess possible portfolio losses in the late 1990s (Linsmeier and Pearson, 1996; Larsen et al., 2002; Jorion, 2007): $VaR$ is a measure of possible portfolio losses; $TVaR$ is the arithmetic mean of the $VaR$s in the tail of the loss distribution; see (Kaas et al., 2009, Chapter 5) for further details.

The paper is organized as follows. In Section 2 we present a brief introduction to the collective risk model. Section 3 introduces our proposed model and its particular cases, as well as the inference procedure. Section 4 presents an application to a health insurance dataset, where the models described in Section 3 are also compared using different model criteria. Section 5 presents a simulation study to evaluate the efficiency of the estimation procedure. Finally, Section 6 provides some conclusions and suggestions for further research.

## 2. Hierarchical collective risk model: a brief review

The basic collective risk model comes from important contributions of Harald Cramér and Filip Lundberg to insurance risk theory. Therefore, it is also known as the Cramér-Lundberg model; see Embrechts et al. (1997, Chapter 1) for details. Migon and Moura (2005) considered a hierarchical extension of the Cramér-Lundberg model and applied it to a health insurance plan. They predicted the pure premium based on the available past information concerning the number and total amount of claims, and also the population exposed to risk.

Let $N_{a,t}$ and $X_{a,t}$ respectively be defined as the number of claims and the total value of claims produced by a portfolio of policies in a given time $t = 1, \cdots, T$ for the age class $a = 1, \ldots, A$.

$$X_{a,t} = \begin{cases} \sum_{j=1}^{N_{a,t}} Z_{a,t,j}, & \text{if } N_{a,t} > 0, \\ 0, & \text{in otherwise.} \end{cases}$$

The $Z_{a,t,j}$ term is the value of the $j^{th}$ claim at time $t$ for the age class $a$.

The main assumptions in the Cramér-Lundberg process are:

i. the number of claims in the interval $(t-1, t)$ is a random variable denoted by $N_{a,t}$; conditional on $N_{a,t} = n_{a,t}$, the claim sizes $Z_{a,j}, j = 1,2,\cdots, n_{a,t}$, are positive independent and identically distributed random variables with finite mean $\mu_a = E(Z_{a,j})$ and variance $\sigma_a^2 = var(Z_{j,a}) < \infty$; and

ii. the claim timing occurs at random instants $t_{1,a} \leq t_{2,a} \leq \cdots$ and the inter-arrival times $T_{k,a} = t_{k,a} - t_{k-1,a}$ are assumed to be independent and identically exponentially distributed random variables with finite mean $E(T_{k,a}) = \lambda_a^{-1}$.

Assuming that the sequences $T_k$ and $Z_k$ are independent of each other and the above conditions hold, it follows that $N_{a,t}$ is a homogeneous Poisson process with rate $\lambda_a$.

If a general distribution with finite mean $\lambda_a^{-1}$ is assumed for the inter-arrival time, a more general process is introduced, known as the renewal model, which allows for a slightly more general process $N_{a,t}$.

Supposing that $Z_{a,t,j} \sim \mathcal{G}(\kappa_a, \theta_a)$ and that the inter-arrival times are exponentially distributed, Migon and Moura (2005) concluded that the claims follow a Poisson process, and the claim sizes are independent and identically gamma distributed:

$$N_{a,t} \mid \lambda_a, \pi_{a,t} \sim \mathcal{P}(\lambda_a \pi_{a,t}), \quad \lambda_a > 0$$

$$X_{a,t} \mid n_{a,t}, \theta_a \sim \mathcal{G}(\kappa_{a,t}, \theta_a), \quad \theta_a > 0 \qquad (1)$$

where the symbol $Y \sim \mathcal{G}(a, b)$ generically denotes that $Y$ is gamma distributed with a probability density function given by $p(y \mid a, b) = \frac{b^a}{\Gamma(a)} y^{a-1} \exp(-by)$; $n_{a,t}$ is the observed number of claims; $\kappa_{a,t} = n_{a,t} \kappa_a$, and $\pi_{a,t}$ denotes the insured population at time $t$ for the age class $a$.

Migon and Moura (2005) modeled the evolution of the insured population via a generalized hierarchical exponential growth model. In particular, we consider the case where the insured population at each time $t$ for age class $a$ is known. This implies that our "Hierarchical Collective Risk model" (HCR) does not assume exchangeability by age of class like the baseline model. The hierarchical modeling comes from the nested probabilistic structure given by the conditional distribution of the total value of claims $X$ given the number of claims $N$. To highlight this characteristic, the directed acyclic graph (DAG) is shown in Figure 1.

## 3. Heavy-tailed and Overdispersed Hierarchical Collective Risk Model

This section presents our model, which is suitable to accommodate high claim values. Its flexibility also enables taking into account possible overdispersion of the claim number.

Figure 1 presents the DAG related to our proposed model. Square nodes denote known constants while circular nodes describe stochastic quantities. The stochastic or functional conditional dependences are denoted by arrows. Two nested round-edged rectangles define the hierarchy of the model. The inner one with random variable $N$ shows the count sub-model and the outer one indicates the conditional dependence of X with respect to $N$ ($X|n$).

Figure 1 Directed acyclic graph (DAG) of the proposal model

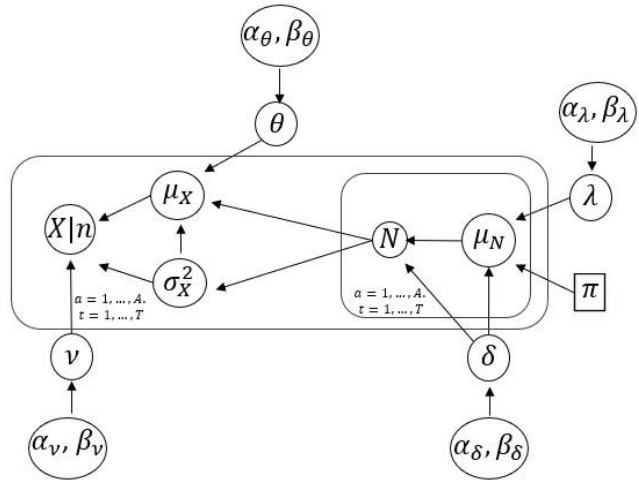

### 3.1 Model specification

It is assumed that the claim value, $X_{a,t}$, $\forall a$ and $\forall t$ is log Student-t ($\mathcal{LT}$) distributed, $X_{a,t} \sim \mathcal{LT}\left(\mu_{a,t}^X, \sigma_{a,t}^2, \nu_a\right)$, i.e., the density function of $X_{a,t}$ is given by:

$$p(x_{a,t} \mid \mu_{a,t}^X, \sigma_{a,t}^2, \nu_a) = \frac{\gamma\left[\frac{(\nu_a+1)}{2}\right]}{\gamma\left(\frac{\nu_a}{2}\right)\sqrt{\pi \nu_a \sigma_{a,t}^2} x_{a,t}} \left[1 + \frac{(\log x_{a,t} - \mu_{a,t})^2}{\nu_a \sigma_{a,t}^2}\right]^{-\frac{(\nu_a+1)}{2}},$$

$$0 \leq x_{a,t} < \infty, \quad \nu_a > 0, \qquad (2)$$

where $\mu_{a,t}^X$, $\sigma_{a,t}^2$ and $\nu_a$ denote the parameters of location, scale and degrees of freedom of the $\mathcal{LT}$ distribution, respectively. We also assume conditional independence for all time periods $t$ and all age risk classes $a$, and we set $\kappa_a = 1, \forall a = 1, \ldots, A$. To facilitate the comparison with the model stated in (1), we set:

$$\mu_{a,t}^X = \log\left(\frac{n_{a,t}}{\theta_a}\right) - \frac{1}{2}\sigma_{a,t}^2, \quad \text{and} \quad \sigma_{a,t}^2 = \log\left(\frac{1}{n_{a,t}} + 1\right). \qquad (3)$$

The equations in (3) are obtained by respectively equating the mean and variance of the $\mathcal{LT}$ distribution in (2) with the mean and variance of the gamma distribution in (1), where $\kappa_a = 1, \forall a = 1, \ldots, A$.

Additionally, the second hierarchical level, related to the overdispersion issue, is dealt with by replacing the Poisson distribution ($\mathcal{P}$) presented in Section 2 with the negative binomial distribution ($\mathcal{NB}$), i.e., $N_{a,t} \mid \mu_{a,t}^N, \delta_a \sim \mathcal{NB}\left(\left(1 + \frac{\mu_{a,t}^N}{\delta_a}\right)^{-1}, \delta_a\right)$, with mean $\mathcal{M} = \mu_{a,t}^N / \delta_a$ and variance $\mathcal{M}(1 + \mathcal{M})$, where $\mu_{a,t}^N = \lambda_a \pi_{a,t}$. This model is parameterized to enable comparisons with the model presented in Section 3. Note that $N_{a,t} \mid \mu_{a,t}, \delta_a$ can be obtained as a mixture of $N_{a,t} \mid \mu_{a,t}, \gamma_a \sim \mathcal{P}(\mu_{a,t}\gamma_a)$ and $\gamma_a \sim \mathcal{G}(\delta_a, \delta_a)$.

To complete the model specification, we assign a proper prior distribution to the model parameters. Independent priors of the model parameters across age class are employed. The choice of the prior distributions was driven by the need to make inferences with minimum prior subject information. For the degrees of freedom parameter, $\nu_a$, a gamma distribution, $\mathcal{G}(a_\nu, b_\nu)$, is used (Fernandez and Steel (1999)) to specify priors with exponential tails. We assume $\delta_a \sim \mathcal{G}(a_\delta, b_\delta)$ for the dispersion parameter. The conditional distributions of the $\theta_a$ and $\lambda_a$ parameters are assigned as $\theta_a \sim \mathcal{G}(a_\theta, b_\theta)$ and $\lambda_a \sim \mathcal{G}(a_\lambda, b_\lambda)$, respectively. To allow subjectivity in the choice of the hyperparameter, a third hierarchical level assigns to all of them $\mathcal{G}(0.1, 0.1)$ and Half-Cauchy $\mathcal{HC}(0,1)$ independent prior distributions (Polson and Scott, 2012; Klein et. al., 2014; Klein and Kneib, 2016).

A sensitivity analysis verified that the assumption of a gamma prior for the hyperparameters does not show anomalous behavior (see Appendix) when used to calculate the 95th percentile of the marginal posterior distribution of the premium.

Table 1 presents a summary of the six models analyzed in this work.

Table1: Distributions of the claim values and the number of claims

| Model | Claim value | Number of claims |
|---|---|---|
| | $X_{a,t} \mid \mu^X_{a,t}, \sigma^2_{a,t}, \nu_a$ | $N_{a,t} \mid \pi_{a,t}, \lambda_a, \delta_a$ |
| (M1) $\mathcal{G} - \mathcal{P}$ | $\mathcal{G}(n_{a,t}, \theta_a)$ | |
| (M2) $\mathcal{LN} - \mathcal{P}$ | $\mathcal{LN}(\mu^X_{a,t}, \sigma^2_{a,t})$ | $\mathcal{P}(\lambda_a \pi_{a,t})$ |
| (M3) $\mathcal{LT} - \mathcal{P}$ | $\mathcal{LT}(\mu^X_{a,t}, \sigma^2_{a,t}, \nu_a)$ | |
| (M4) $\mathcal{G} - \mathcal{NB}$ | $\mathcal{G}(n_{a,t}, \theta_a)$ | |
| (M5) $\mathcal{LN} - \mathcal{NB}$ | $\mathcal{LN}(\mu^X_{a,t}, \sigma^2_{a,t})$ | $\mathcal{NB}\left(\left(1+\frac{\mu^N_{a,t}}{\delta_a}\right)^{-1}, \delta_a\right)$ |
| (M6) $\mathcal{LT} - \mathcal{NB}$ | $\mathcal{LT}(\mu^X_{a,t}, \sigma^2_{a,t}, \nu_a)$ | |

## 3.2 Inference procedure

Let $\Theta = (\theta, \kappa, \nu, \lambda, \delta)$ be the vector of parameters for all risk classes a, a = 1, ..., A and denote by **D** all information available for all time periods t, t = 1, ..., T. Therefore, the likelihood function of the $\mathcal{LT} - \mathcal{NB}$ model is given by:

$$\log L(\Theta \mid D) = \sum_{a,t}^{A,T} \log \frac{\nu_a^{\frac{\nu_a}{2}} w(\nu_a)}{\log(1 + n_{a,t})^{\frac{1}{2}}} - \frac{(\nu_a + 1)}{2} \log(\nu_a + P_{a,t}^2)$$
$$+ \log \psi_a \tag{4}$$

with $w(\nu_a) = \frac{\gamma\left[\frac{(\nu_a+1)}{2}\right]}{\gamma\left(\frac{\nu_a}{2}\right)\gamma\left(\frac{1}{2}\right)}$, $P_{a,t} = \sigma_{x_{a,t}}^{-1} \left[\log\left(\frac{x_{a,t}\theta_a(1+n_{a,t})^{\frac{1}{2}}}{(n_{a,t})^{\frac{3}{2}}}\right)\right]$ and $\psi_a = \frac{\Gamma(n_{a,t}+\delta_a)}{\Gamma(n_{a,t})\Gamma(\delta_a)} \frac{\left(\frac{\mu_{a,t}}{\delta_a}\right)^{n_{a,t}}}{\left(1+\frac{\mu_{a,t}}{\delta_a}\right)^{\delta_a+n_{a,t}}}$.

The inference procedure is carried out by sampling from the posterior distribution. The posterior distribution $\pi^p(\Theta \mid \mathcal{D})$ is proportional to the likelihood $L(\Theta \mid D)$ times the prior information $\pi(\Theta)$. Sampling from $\pi^p(\Theta \mid \mathcal{D})$ can be done by the Markov chain Monte Carlo (MCMC) technique.

## 3.3 The premium value

Migon and Moura (2005) obtained the premium value as a solution to a decision problem. Using the modified absolute deviation utility with some appropriate constants, it can be proved under some regularity conditions that the value $d$ that maximizes the utility function is the $95^{th}$ percentile of the predictive posterior of the unknown total claim value per insured.

Let the total claim value in the time horizon and the insured population for each age class $a$, $a = 1, ..., A$, at time $T + H$, respectively, be $X_{a,T+H}$ and $\Pi_{a,T+H}$, where $H$ is the planning horizon. Based on the decision theory fundamentals, if the modified absolute deviation is used as a utility function, then the decision value (premium value) that maximizes the utility function in the marginal predictive posterior distribution ($R_{a,T+H} = \frac{X_{a,T+H}}{\Pi_{a,T+H}}$) is given by its $95^{th}$ percentile. Therefore, the premium value is obtained for each model as the $95^{th}$ percentile of $R_{a,T+H}$, which is calculated by:

$$p(R_{a,T+H} \mid D_T) = \int_\Theta p(R_{a,T+H} \mid \theta) p(\theta \mid D_T) d\Theta,$$

where $p(\theta \mid D_T)$ denotes the posterior distribution of all model parameters.

### 3.4 Measure of risk in the marginal predictive posterior distribution

The marginal predictive distribution of R (see Section 3) is used to obtain the Value at Risk ($VaR$) and the Tail Value at Risk ($TVaR$).

Let R be a random variable that represents the risk. The Value at Risk of $R$ at a level $\tau \in (0,1)$, $\text{VaR}_R(\tau)$, is defined as $VaR_R(\tau) \overset{(def)}{=} F_R^{-1}(\tau) \overset{(not)}{=} \inf\{r : F_R(r) \geq \tau\}$. The $VaR$ measure is the inverse cdf of R computed at $\tau$. The Tail Value at Risk of $R$ at level $\tau$ is defined as $\text{TVaR}_R(\tau) = \frac{1}{1-\tau} \int_\tau^1 \text{VaR}_R(t) dt$. Therefore, $TVaR$ can be interpreted as the arithmetic mean of the $VaR$s of $R$ from $\tau$ onward. An alternative expression is given by $\text{TVaR}_R(\tau) = \text{VaR}_R(\tau) + \frac{1}{1-\tau} \text{ES}_R(\tau)$, where $\text{ES}_R(\tau)$ describes the expected shortfall (ES) at level $\tau$ of $R$. The ES is defined as $\text{ES}_R(\tau) = E\left[(R - \text{VaR}_R(\tau))_+\right]$, and ES can be interpreted as the net stop-loss premium in the hypothetical situation of excess over VaR. The second form of $TVaR$ is used in Section 4.

### 3.5 Bayesian model selection

To assess model fitness, the deviance information criterion (DIC) (Spiegelhalter et al., 2002) is used. The DIC is defined by $DIC = D(\overline{\Theta}) + 2p_D$, where $D(\Theta) = -2\log(L(\Theta \mid y))$, with y denoting all information provided by the data, $\Theta$ is the vector of parameters and $L(\Theta \mid y)$ is the likelihood function, $p_D = \overline{D(\Theta)} - D(\overline{\Theta})$, where $\overline{D(\Theta)}$ and $\overline{\Theta}$ denote the posterior mean of $D(\Theta)$ and $\Theta$, respectively. To verify the accuracy of prediction values, the continuous ranked probability score (CRPS) criterion (Lopes et al., 2012) was also applied. The CRPS can be expressed as:

$$CRPS = \sum_{a,t}^{A,T} \left( E|y_{rep,at} - y_{at}| - 0.5 E|y_{rep,at} - \tilde{y}_{rep,at}| \right)/AT,$$

where the expectations in (5) are calculated under the posterior marginal distribution, and $y_{rep,at}$ and $\tilde{y}_{rep,at}$ are indepependent replicates from it. Assuming there is a sample of size L of the marginal posterior distribution, these expectations can be approximated by $\sum_{l=1}^{L} |y_{rep,at}^{(l)} - y_{at}|/L$ and $\sum_{l=1}^{L} |y_{rep,at}^{(l)} - \tilde{y}_{rep,at}^{(l)}|/L$, respectively. Models with lower DIC and CRPS values are preferred.

### 4. Illustration with a real dataset

The dataset used to illustrate our alternative models was extracted from a health care plan implemented in a university in northeast Brazil. This dataset was also used by Migon and Moura (2005). Its main characteristics are being self-administered and having a small number of insureds. The dataset consists of only 12 monthly observations of the number of claims, the total value of the observed claims, and the number of insured persons per type of service (doctor's office visit, diagnostic test, and hospital care), by age class. There are seven age classes, $a = 1, ..., 7$ and they respectively correspond to the following intervals:[0,17), [17,30), [30,40), [40,50), [50,60), [60,70), [70 +). For the sake of simplicity, the notation employed in the next sections does not include an additional index to describe the type of service. All calculations were performed separately for each type of service, so the respective predicted values of the size and the number of claims per age class were obtained by aggregation.

The six models described in Table 1 of Section (3) were fitted to the data. Sampling from $\pi^p(\Theta \mid \mathcal{D})$ is done by the Markov chain Monte Carlo (MCMC) method, using the No-U-Turn-Sampler implemented in the Stan software (Team, 2018). Three chains of length 10,000, starting from different locations with burn-in of 5,000 iterations, were generated. Convergence was monitored via Markov chain Monte Carlo: autocorrelation; cross-correlation; and density plots using the approach developed by Cowles and Carlin (1996) and implemented in R (R Development Core Team, 2018).

The marginal predictive posterior distribution $R_{a,T+H}$ for the considered models is presented in Figure 1(a), where (+) corresponds to the $95^{th}$ percentile of the predictive distribution, accumulated over the planning horizon.

Figure 2 Boxplot of the marginal predictive posterior distribution $R_{a,T+H}$ based on the six models by age class.

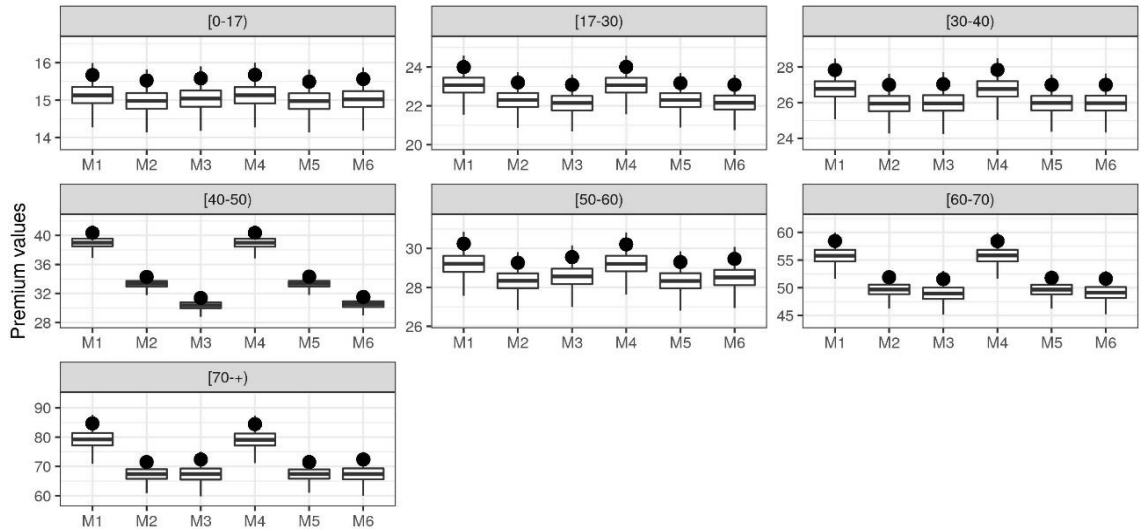

Age classes: $[0,17)$, $[17,30)$, $[30,40)$, $[40,50)$, $[50,60)$, $[60,70)$, $[70,+)$. Models M1 ($\mathcal{G}-\mathcal{P}$), M2 ($\mathcal{LN}-\mathcal{P}$), M3 ($\mathcal{LT}-\mathcal{P}$), M4 ($\mathcal{G}-\mathcal{NB}$), M5 ($\mathcal{LN}-\mathcal{NB}$), M6 ($\mathcal{LT}-\mathcal{NB}$). The solid circle symbol indicates the $95^{th}$ percentile.

These results indicate that the premiums obtained from the $\mathcal{LT}$ models are considerably smaller for age class $[40,50)$ than the ones obtained from the other models. Thus, the $\mathcal{LT}$ models provide greater shrinkage of the premium. For the age classes $[60,70)$ and $[70,+)$, the marginal posteriors of $R_{a,T+H}$ are approximately the same for the $\mathcal{LN}-\mathcal{NB}$ and $\mathcal{LT}-\mathcal{NB}$ models, but different for the $\mathcal{G}-\mathcal{NB}$ model.

We additionally considered an analysis by age class of the posterior distributions of δ and ν, shown in Figure 2. An analysis of the posterior distribution of **δ** shows there is a high probability of the true values of **δ** being greater than 1 for all three considered models and all age classes. The small values of the **ν** estimates indicate heavy tails, most evident for the $[40,50)$ age class. These results show evidence of overdispersion and the need to introduce a heavy-tailed distribution to model the claim values.

Figure 3 shows box-plots of the marginal posterior of the (a) **λ** and (b) **θ** effects based on the six models by age class. As expected, an analysis by age class of the posterior distributions of **λ** for all six models shows their posterior means increase as the insured gets older. Nevertheless, the three models that take into account overdispersion have more uncertainty in estimating **λ** than their respective counterparts. The parameter θ$_a$ increases as the respective mean claim value decreases (see equation 3). Since the mean claim value tends to increase with age, the pattern shown in Figure 3(b) is expected. It is worth noting that the $\mathcal{LT} - \mathcal{NB}$ and $\mathcal{LT} - \mathcal{P}$ models perform more in line with this pattern than the other models.

Figure 3. Box-plots of the marginal posterior distribution of the (a) $\delta$ and (b) $\nu$ effects based on six models by age class.

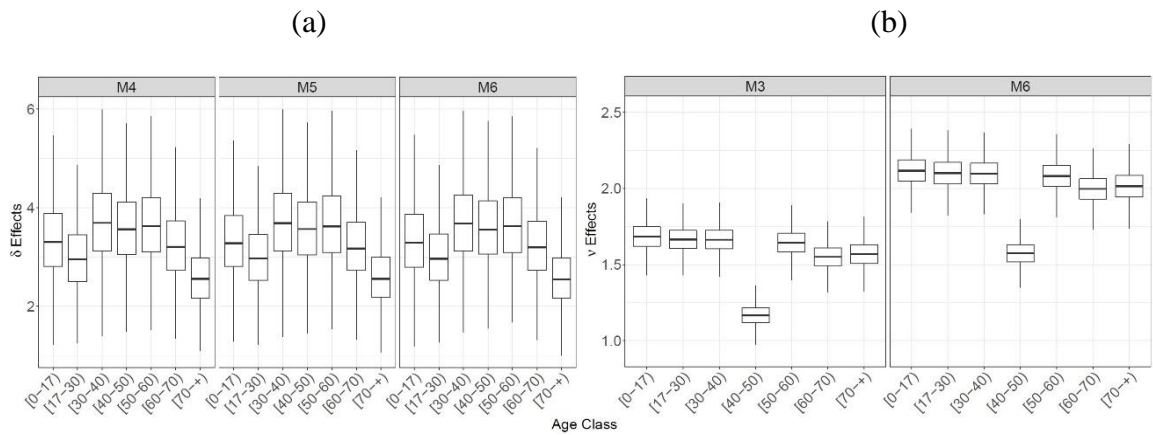

Models M1 ($\mathcal{G} - \mathcal{P}$), M2 ($\mathcal{LN} - \mathcal{P}$), M3 ($\mathcal{LT} - \mathcal{P}$), M4 ($\mathcal{G} - \mathcal{NB}$), M5 ($\mathcal{LN} - \mathcal{NB}$), M6 ($\mathcal{LT} - \mathcal{NB}$)}

Figure 4. Box-plots of the marginal posterior distribution of the (a) λ and (b) θ effects based on six models by age class.

(a)                                    (b)

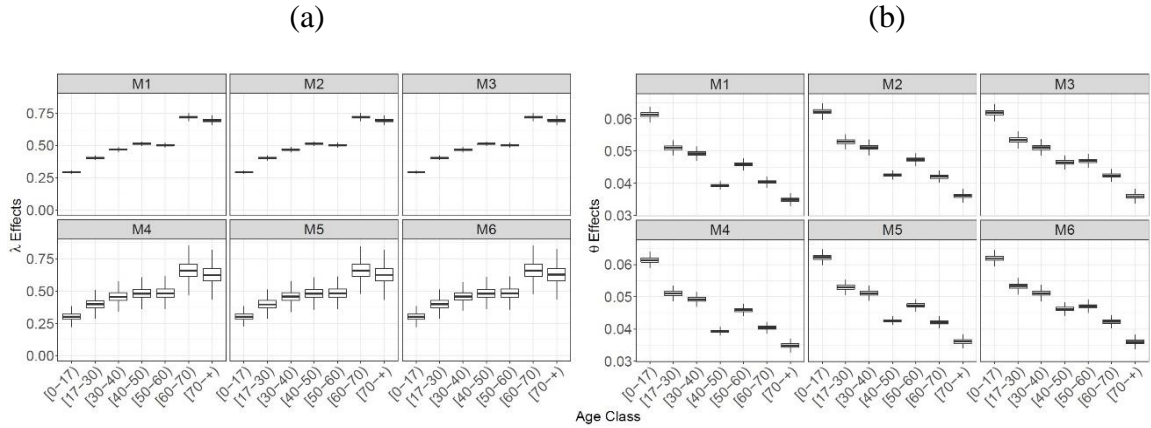

Models M1 ($\mathcal{G}-\mathcal{P}$), M2 ($\mathcal{LN}-\mathcal{P}$), M3 ($\mathcal{LT}-\mathcal{P}$), M4 ($\mathcal{G}-\mathcal{NB}$), M5 ($\mathcal{LN}-\mathcal{NB}$), M6 ($\mathcal{LT}-\mathcal{NB}$)}

Some measurements to study the impact of overdispersion on the age classes were also developed. These measurements are based on the posterior predictive variance of both size and number of claims. To quantify the over or underdispersion, the variance of the posterior marginal predictive distribution of $N_{a,t}$ is calculated. The term $1 + \frac{\mu_{a,t}^N}{\delta_a}$ measures the excess of variability in each age class a.

Figure 4 shows the coefficient of variation by age class (a); and for the age class and the three types of service (circle, triangle, square) (b), for model M6. The hospital care service (square) presents the largest premium values. The age classes $[40-50)$ and $[70,+)$ show the largest coefficients of variation when all the services are averaged.

Figure 5 Marginal posterior distributions of the variation coefficients based on M6: $\mathcal{LT} - \mathcal{NB}$ by: (a) Age class and service, each curve corresponds to T = 12 values. (b) Age class, each curve corresponds to T × S = 36 values.

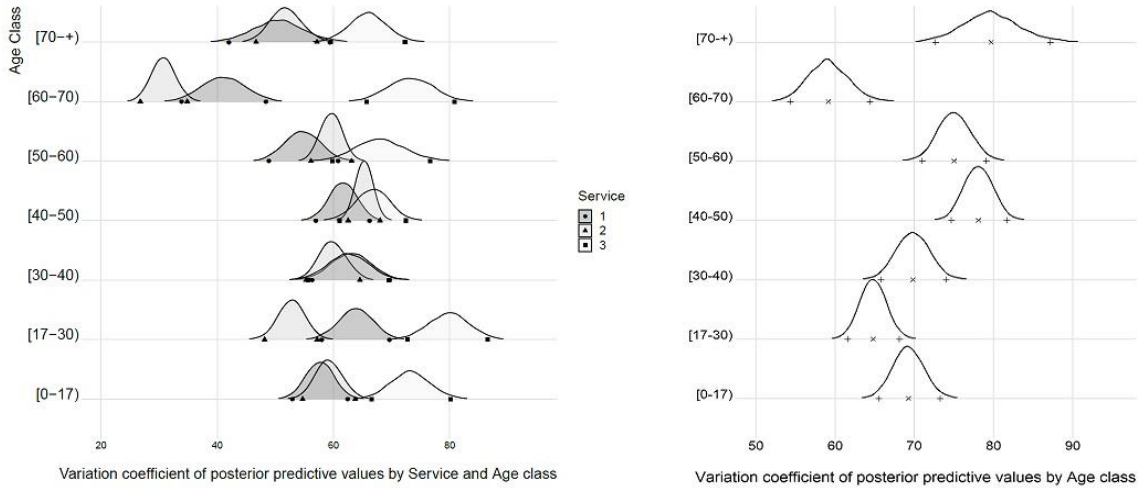

Notes: In (a) the solid circle symbol and gray curve indicate information with respect to service 1. The solid triangle symbol and light gray curve indicate information with respect to service 2. The solid square symbol and white curve indicate information about service 3. The distance between solid symbols (circle, triangle, square) denote the 2.5%, 97.5% empirical quantiles with respect to services 1, 2 and 3, respectively, for each age class. In (b), the symbol (+) denotes the 2.5%, 97.5% empirical quantiles and × the posterior mean by age class without services.

To identify possible high-risk age groups assuming the presence of overdispersion, that is, assuming a negative binomial distribution for the number of claims, the measures VaR and TVaR were calculated for the models whose number of claims were assumed to be negative binomial distributed, i.e., models M4, M5 and M6. These calculations were obtained for each age class; see Figures 5 and 6, respectively.

Figure 5 (a) shows that all age classes provide high VaR when the distribution of the value of claims is associated with the gamma distribution. Conversely, the Student $\mathcal{LT}$-distribution for the value of claims is associated with low values of $VaR$. The $VaR$ values for age class $[0,17)$ are relatively close for all models. For the age classes $([40,50), [60,70), [70, +))$, the values of $VaR$ are considerably different when compared to the other age classes. Furthermore, the $VaR$ estimates from $\mathcal{LN}$ distribution is very close to those obtained from the Student $\mathcal{LT}$ distribution, except for the age class[40,50). Figure 5 (b) describes the tail of the value at risk in τ ∈ (0,0.1).

Figure 6 Values-at-risk calculated for the seven age classes and for models M4, M5 and M6: (a) VaR calculated over a complete range of confidence levels τ ∈ (0, 1). (b) $VaR$ calculated over a range of confidence levels τ ∈ (0, 0.1).

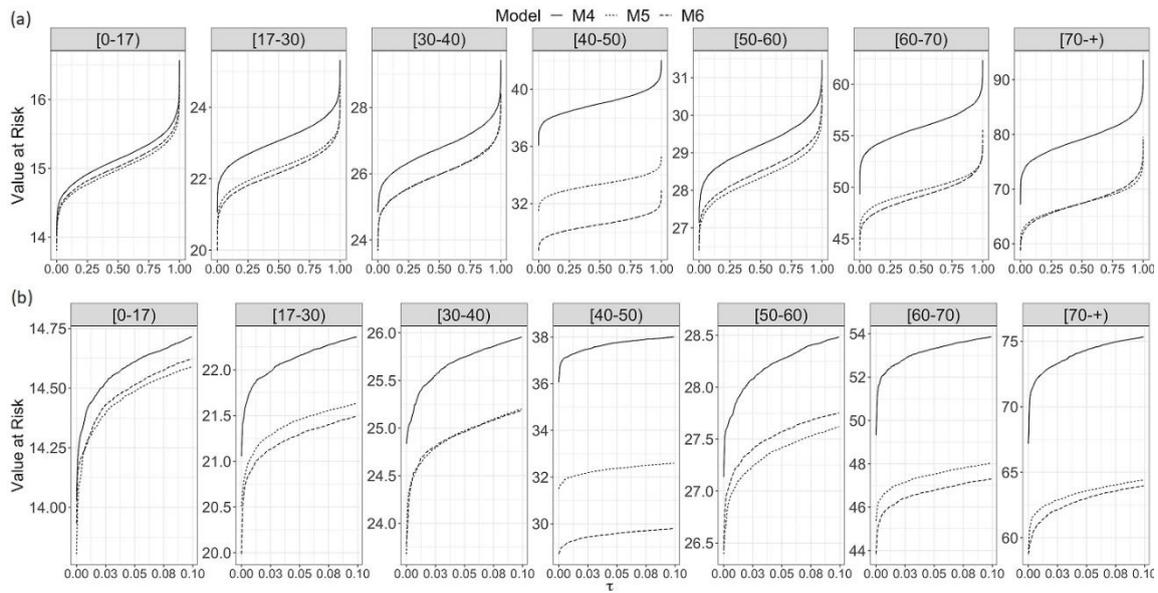

Notes: Models M4 ($\mathcal{G} - \mathcal{NB}$), M5 ($\mathcal{LN} - \mathcal{NB}$), M6 ($\mathcal{LT} - \mathcal{NB}$)}. Quantile τ.

According to Figure 6, TVaRstatistics presents approximately the same values for age classes [0,17), [17,30), [30,40), [50,60). Higher values of $TVaR$ are observed for model M4 for all age classes.

Figure 7. Tail Values-at-risk calculated for the seven age classes and for models M4, M5 and M6. (a) TVaR calculated over a complete range of confidence levels $\tau \in (0, 1)$. (b) $TVaR$ calculated over a range of confidence levels $\tau \in (0, 0.1)$

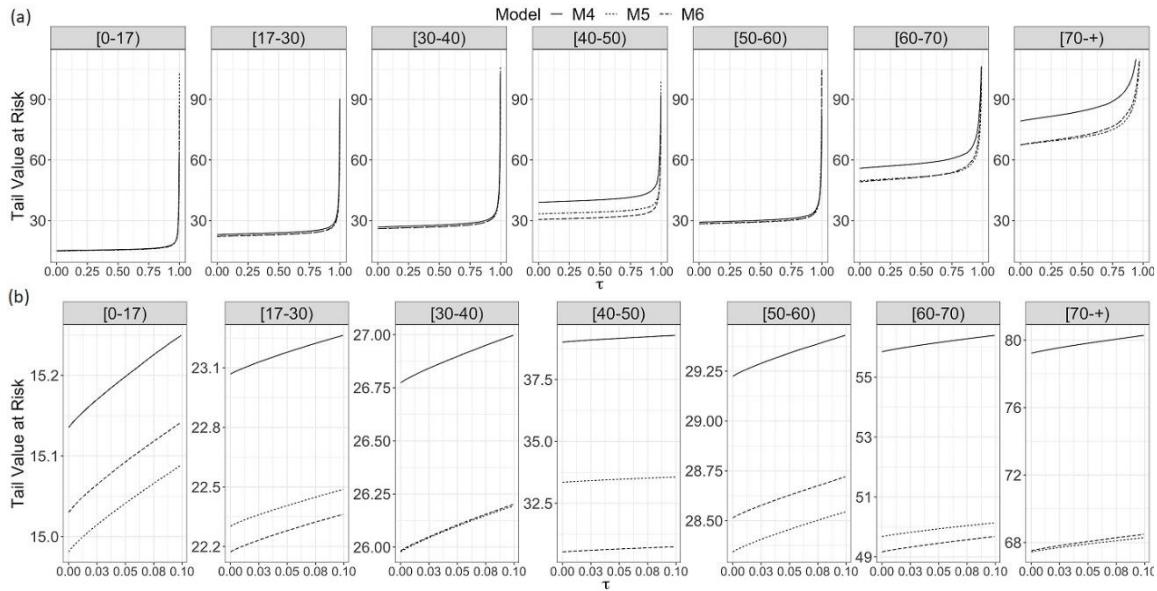

Notes: Models M4 ($\mathcal{G} - \mathcal{NB}$), M5 ($\mathcal{LN} - \mathcal{NB}$), M6 ($\mathcal{LT} - \mathcal{NB}$)}. Quantile $\tau$.

Table 2 presents the values of model comparison measures. According to these criteria, the $\mathcal{LT} - \mathcal{NB}$ distributions seem to be the most appropriate to model the number of claims and their total value.

Table 2 Model comparison criteria for the six models.

| Model | Comparison criterion | | | |
|---|---|---|---|---|
| | $\overline{D(\Theta)}$ | $D(\overline{\Theta})$ | DIC | CRPS |
| M1: $\mathcal{G} - \mathcal{P}$ | 10420.140 | 10378.410 | 10461.860 | 0.491 |
| M2: $\mathcal{LN} - \mathcal{P}$ | 8172.317 | 8130.900 | 8213.733 | 0.465 |
| M3: $\mathcal{LT} - \mathcal{P}$ | 6077.799 | 6036.067 | 6119.53 | 0.454 |
| M4: $\mathcal{G} - \mathcal{NB}$ | 8761.463 | 8582.174 | 8940.752 | 0.581 |
| M5: $\mathcal{LN} - \mathcal{NB}$ | 7349.903 | 6125.228 | 8574.577 | 0.646 |
| M6: $\mathcal{LT} - \mathcal{NB}$ | **4005.113** | **3957.580** | **4052.646** | **0.399** |

The numbers in boldface are the smallest. DIC: Deviance Information Criterion, CRPS: Continuous Ranked Probability Score.

## 5. A simulation study

In this section we present a simulation study to evaluate the efficiency of the estimation procedure in recovering the true values of the model hyperparameters $(\alpha_\lambda, \beta_\lambda, \alpha_\delta, \beta_\delta, \alpha_\theta, \beta_\theta, \alpha_\nu, \beta_\nu)$. Three hundred datasets based on M6 were generated and the population size was kept fixed and taken from the health insurance dataset described in Section 4.

To make the simulated data as similar as possible to the real data, the true hyperparameters in Table 3 were calculated by equating the corresponding means of the gamma distribution to the posterior means obtained by fitting the real data and setting their respective precision values approximately equal to 11. The system of two equations was solved and the respective true values were obtained.

The prior distributions used to fit the $S = 300$ simulated datasets were $\mathcal{G}(0.1, 0.1)$ for the eight hyperparameters.

To obtain samples of the posterior distribution, the MCMC algorithm was implemented using the Stan software (Team, 2018). To monitor the convergence of chains, the procedures described in Section 3.2 were implemented.

Table 3 Estimated values to evaluate the performance of the MCMC results based on the 300 simulated datasets

| Values | Hyperparameters | | | | | | | |
|---|---|---|---|---|---|---|---|---|
| | $\alpha_\lambda$ | $\beta_\lambda$ | $\alpha_\delta$ | $\beta_\delta$ | $\alpha_\theta$ | $\beta_\theta$ | $\alpha_\nu$ | $\beta_\nu$ |
| True | 2.85 | 5.62 | 12.37 | 3.71 | 1.11 | 11.11 | 10.12 | 3.35 |
| MRB (in%) | 1.75 | 0.41 | 0.58 | 0.02 | -0.30 | 0.13 | 0.23 | -1.69 |
| MSE (in%) | 1.63 | 0.36 | 0.65 | 1.38 | 2.73 | 0.06 | 0.17 | 1.05 |

Note: Mean relative bias (MRB) and mean squared error (MSE)

The mean relative bias (MRB), in percentage, is defined as $\text{MRB} = 100 \cdot \theta^{-1} \frac{\sum_{s=1}^{S}\left(\widehat{\theta^{(s)}} - \theta\right)}{S}$, where $\widehat{\theta^{(s)}}$ is the posterior mean of the respective true hyperparameter $\theta$ for the $s$ simulated datasets, $s = 1, \ldots, S$. Another criterion to evaluate the performance of the MCMC results is given by the mean squared error, expressed in Table 3 in percentage, i.e., $\text{MSE} = 100 \cdot \frac{\sum_{s=1}^{S}\left(\widehat{\theta^{(s)}} - \theta\right)^2}{S}$. Table 3 shows small MRB and MSE values.

Since the premium value P is a conservative measurement (see Migon and Moura, 2005), we developed a statistic $U_a$ that compares the simulated value of $R_a$, with the 95% quantile obtained from the marginal posterior predictive distribution of $R_a$, i.e., the premium value $P_a$; see Section 3.3 for details. The quantity $U_a$ indicates the level of protection that the proposed model gives to the insurance company for each age class $a = 1, \ldots, A$.

Table 4 Percentage of cases where the estimated quantile was higher than the threshold (premium calculated using the simulated dataset by age class)

| Statistic | Age class | | | | | | |
|---|---|---|---|---|---|---|---|
| | 0 ⊢ 17 | 17 ⊢ 30 | 30 ⊢ 40 | 40 ⊢ 50 | 50 ⊢ 60 | 60 ⊢ 70 | ≥ 70 |
| $U_a$ | 0.68 | 1.03 | 3.77 | 1.71 | 2.05 | 1.37 | 3.08 |

Note: $U_a$ indicates the level of protection

Table 4 shows the values of $U_a = 100 \cdot \sum_{s=1}^{S} \frac{\#Q_s(R_a, 0.95) > P_a^{(s)}}{S}$, where $Q_s(R_a, 0.95)$ denotes the 95% quantile of $R_a$ and $P_a^{(s)}$ defines the true premium value, calculated for each simulated sample $s = 1, \ldots, S$ and age class $a = 1, \ldots, A$. Although the age classes $30 \vdash 40$ and $\geq 70$ yielded slightly higher percentages than the others, on average all values of $U_a$ were approximately 2%, indicating that in only approximately 2% of the samples were the premium values smaller than the total claim value per insured.

## 6. Main Conclusions and final remarks

Collective risk models that incorporate heavy-tailed distributions and overdispersion were proposed to provide realistic estimates and forecasts, allowing practitioners to set fair premium prices.

Insurance premium values not greatly affected by observations far from the center and allowing for the number of claims to have a larger variance than that prescribed by the Poisson distribution were calculated. The inference procedure was applied in the Bayesian approach and estimation was carried out through MCMC methods via the Stan software (Team, 2018). Table 2 shows comparison criterion values in favor of our proposal.

An application was presented and six hierarchically structured models (Gelman and Hill, 2007) were fitted. The results showed shrinkage of the premium for some age classes. Excess of variability and variation coefficients were used to identify age classes with non-standard premiums. The analysis of the insurance data revealed that service **3** (hospitalization) had more observations far from the center than the other services. The estimates of the parameters δ and ν presented in Figure 2 suggest overdispersion and heavy tails.

The VaR and TVaR measures were considerably different for the three models analyzed. This was due to the difference in the fat tails of models M4, M5 and M6. This result is consistent with the findings of Stoyanov et al. (2013) and Günay (2017). Furthermore, in the alternative formula for TVaR, the expectation shortfall term was taken into account for the premium calculation. This provides greater flexibility in the interpretation. The method shows evidence that the age classes [40,50), [60,70) and [70, +) have higher risk than the others (see Figures 1, 5 and 6).

To validate the algorithm, artificial data were generated from the proposed model ($M6$). The convergence of the chains of the parameters was verified by Geweke and Raftery's method. The model correctly predicted the quantities of interest. The inference procedure showed the desired behavior, since the premium calculated via our proposed model did not cause a financial loss to the insurance company compared to the results of the standard model (see Figure 1). Furthermore, when we compared the most important quantity to be considered, the premium value, it was suitable, as indicated by the $U_a$ statistic.

For future work, a model that relates the costs and the age of the beneficiaries could be considered. The Birnbaum Saunders-T and inverse Gaussian distributions should also be evaluated and used to model the claim values (Gilberto et al., 2012). Reference priors for the overdispersion parameter based on Liseo et al. (2010) could also be obtained and applied.

Covariates such as gender, education level and national economic indices could be included in the model to help to obtain the fairest premium values.

**Appendix**

Premium value for each model by age class using gamma ($\mathcal{G}$) and half-Cauchy ($\mathcal{HC}$) distributions for the hyperparameters.

|  |  | [0 − 17) | | [17 − 30) | | [30 − 40) | | [40 − 50) | | [50 − 60) | | [60 − 70) | | [70 − +) | |
|---|---|---|---|---|---|---|---|---|---|---|---|---|---|---|---|
|  |  | $\mathcal{G}$ | $\mathcal{HC}$ | $\mathcal{G}$ | $\mathcal{HC}$ | $\mathcal{G}$ | $\mathcal{HC}$ | $\mathcal{G}$ | $\mathcal{HC}$ | $\mathcal{G}$ | $\mathcal{HC}$ | $\mathcal{G}$ | $\mathcal{HC}$ | $\mathcal{G}$ | $\mathcal{HC}$ |
| M1 | $l_i$ | 14.520 | 14.530 | 21.954 | 22.014 | 25.570 | 25.576 | 37.551 | 37.476 | 28.059 | 28.072 | 52.906 | 52.981 | 73.326 | 73.295 |
|  | $l_u$ | 15.781 | 15.769 | 24.160 | 24.148 | 28.018 | 28.011 | 40.645 | 40.550 | 30.437 | 30.410 | 58.935 | 59.079 | 85.704 | 85.388 |
|  | $P$ | 15.672 | 15.661 | 23.993 | 23.954 | 27.827 | 27.814 | 40.350 | 40.306 | 30.249 | 30.223 | 58.448 | 58.541 | 84.673 | 84.460 |
| M2 | $l_i$ | 14.382 | 14.374 | 21.277 | 21.270 | 24.802 | 24.863 | 32.217 | 32.193 | 27.238 | 27.322 | 47.241 | 47.348 | 62.834 | 62.936 |
|  | $l_u$ | 15.642 | 15.582 | 23.373 | 23.347 | 27.185 | 27.127 | 34.497 | 34.539 | 29.451 | 29.488 | 52.371 | 52.229 | 72.412 | 72.231 |
|  | $P$ | 15.526 | 15.496 | 23.201 | 23.167 | 27.001 | 26.922 | 34.283 | 34.335 | 29.267 | 29.281 | 51.907 | 51.853 | 71.498 | 71.484 |
| M3 | $l_i$ | 14.423 | 14.426 | 21.086 | 21.111 | 24.783 | 24.807 | 29.219 | 29.246 | 27.391 | 27.416 | 46.175 | 46.344 | 62.055 | 62.009 |
|  | $l_u$ | 15.692 | 15.646 | 23.278 | 23.245 | 27.241 | 27.214 | 31.570 | 31.568 | 29.750 | 29.732 | 52.184 | 52.176 | 73.400 | 73.380 |
|  | $P$ | 15.581 | 15.549 | 23.083 | 23.052 | 27.035 | 27.022 | 31.381 | 31.359 | 29.559 | 29.524 | 51.564 | 51.674 | 72.370 | 72.373 |
| M4 | $l_i$ | 14.523 | 14.536 | 21.980 | 21.990 | 25.552 | 25.560 | 37.481 | 37.564 | 28.079 | 28.110 | 52.843 | 52.831 | 73.310 | 73.289 |
|  | $l_s$ | 15.794 | 15.724 | 24.169 | 24.130 | 28.026 | 27.974 | 40.613 | 40.517 | 30.413 | 30.364 | 58.899 | 59.011 | 85.510 | 85.644 |
|  | $P$ | 15.678 | 15.643 | 24.004 | 23.941 | 27.834 | 27.784 | 40.358 | 40.288 | 30.213 | 30.194 | 58.419 | 58.541 | 84.443 | 84.569 |
| M5 | $l_i$ | 14.394 | 14.371 | 21.287 | 21.298 | 24.797 | 24.840 | 32.192 | 32.181 | 27.243 | 27.235 | 47.150 | 47.417 | 62.880 | 62.863 |
|  | $l_u$ | 15.592 | 15.591 | 23.315 | 23.325 | 27.192 | 27.153 | 34.484 | 34.527 | 29.489 | 29.418 | 52.202 | 52.333 | 72.314 | 72.350 |
|  | $P$ | 15.494 | 15.472 | 23.169 | 23.154 | 26.999 | 26.932 | 34.315 | 34.327 | 29.306 | 29.257 | 51.792 | 51.863 | 71.493 | 71.555 |
| M6 | $l_i$ | 14.430 | 14.420 | 21.132 | 21.119 | 24.806 | 24.813 | 29.428 | 29.424 | 27.404 | 27.388 | 46.423 | 46.467 | 62.125 | 62.334 |
|  | $l_s$ | 15.674 | 15.612 | 23.259 | 23.268 | 27.227 | 27.215 | 31.700 | 31.683 | 29.682 | 29.680 | 52.160 | 52.182 | 73.243 | 73.341 |
|  | $P$ | 15.563 | 15.524 | 23.085 | 23.066 | 26.998 | 27.010 | 31.518 | 31.513 | 29.466 | 29.500 | 51.636 | 51.739 | 72.377 | 72.343 |

Age classes: [0,17), [17,30), [30,40), [40,50), [50,60), [60,70), [70,+). Models M1 ($\mathcal{G} - \mathcal{P}$), M2 ($\mathcal{LN} - \mathcal{P}$), M3 ($\mathcal{LT} - \mathcal{P}$), M4 ($\mathcal{G} - \mathcal{NB}$), M5 ($\mathcal{LN} - \mathcal{NB}$), M6 ($\mathcal{LT} - \mathcal{NB}$). $l_i$ indicates the lower limit's $2.5^{th}$ percentile, $l_u$ indicates the lower limit's $97.5^{th}$ percentile, and P is the estimated premium ($95^{th}$ percentile).